\documentclass[conference]{IEEEtran}
\IEEEoverridecommandlockouts
\usepackage{cite}
\usepackage{amsmath,amssymb,amsfonts}
\usepackage{algorithmic}
\usepackage{graphicx}
\usepackage{textcomp}
\usepackage{xcolor}

\usepackage{multirow}
\usepackage{makecell}
\usepackage{booktabs}
\usepackage{threeparttable}

\usepackage{enumitem}

\usepackage{tikz}
\newcommand*\circled[2]{\tikz[baseline=(char.base)]{
\node[circle,draw,scale=#2,inner sep=2pt] (char) {#1};}}

\def\BibTeX{{\rm B\kern-.05em{\sc i\kern-.025em b}\kern-.08em
    T\kern-.1667em\lower.7ex\hbox{E}\kern-.125emX}}

\begin{document}

\setlength{\intextsep}{0.cm}
\setlength{\textfloatsep}{1.2ex}
\setlength{\floatsep}{1.5ex}
\setlength{\abovecaptionskip}{0.cm}
\setlength{\belowcaptionskip}{0.cm}
\setlength{\dbltextfloatsep}{0.22\baselineskip}

\title{ACCO: Automated Causal CNN Scheduling Optimizer for 
Real-Time Edge Accelerators \\
}

\newcommand{\todo}[1]{{\textcolor{black}{#1}}}
\newcommand{\discussion}[1]{{\textcolor{black}{#1}}}
\newcommand{\mv}[1]{{\textcolor{black}{#1}}}
\newcommand{\lm}[1]{{\textcolor{black}{#1}}}
\newcommand{\mvnok}[1]{{\textcolor{black}{#1}}}
\newcommand{\mvnnok}[1]{{\textcolor{black}{#1}}}
\newcommand{\camreadyok}[1]{{\textcolor{black}{#1}}}

\newcommand{\mvn}[1]{{\textcolor{green}{#1}}}
\newcommand{\mvnn}[1]{{\textcolor{green}{#1}}}
\newcommand{\camready}[1]{{\textcolor{black}{#1}}}

\newcommand\copyrighttext{%
  \footnotesize \textcopyright 2023 IEEE. Personal use of this material is permitted.
  Permission from IEEE must be obtained for all other uses, in any current or future
  media, including reprinting/republishing this material for advertising or promotional
  purposes, creating new collective works, for resale or redistribution to servers or
  lists, or reuse of any copyrighted component of this work in other works.
  DOI: 10.1109/ICCD58817.2023.00065.}
\newcommand\copyrightnotice{%
\begin{tikzpicture}[remember picture,overlay]
\node[anchor=south,yshift=10pt] at (current page.south) {\fbox{\parbox{\dimexpr\textwidth-\fboxsep-\fboxrule\relax}{\copyrighttext}}};
\end{tikzpicture}%
}

\author{
\IEEEauthorblockN{Jun Yin\IEEEauthorrefmark{1}, Linyan Mei\IEEEauthorrefmark{1}, Andre Guntoro\IEEEauthorrefmark{2}, Marian Verhelst\IEEEauthorrefmark{1}}

\IEEEauthorblockA{\textit{\IEEEauthorrefmark{1}ESAT-MICAS KU Leuven, \IEEEauthorrefmark{2}Robert Bosch GmbH}}

\IEEEauthorblockA{\textit{\IEEEauthorrefmark{1}\{jun.yin, linyan.mei, marian.verhelst\}@kuleuven.be \IEEEauthorrefmark{2}andre.guntoro@de.bosch.com}}

\thanks{\camready{This project has been partly funded by the MSCA program under grant agreement No. 956962, the European Research Council (ERC) under grant agreement No. 101088865, the European Union’s Horizon 2020 programme under grant agreement No. 101070374 and the Flanders AI Research Program.}}
\vspace{-3ex}
}

\maketitle

\begin{abstract}
Spatio-Temporal Convolutional Neural Networks \mvnok{(ST-CNN)} allow extending CNN capabilities from image processing to \mvnok{consecutive temporal-pattern recognition}.
Generally, \camreadyok{state-of-the-art (SotA) ST-CNNs inflate the feature maps and weights from well-known CNN backbones to represent the additional time dimension. 
However, \mvnnok{edge computing applications} 
would suffer \mvnnok{tremendously from such large} computation/memory overhead. 
\mvnnok{Fortunately,} the overlapping nature of ST-CNN \mvnnok{enables various} optimization methods, such as the dilated causal convolution structure and Depth-First~(DF) layer fusion to reuse the computation between time steps and CNN sliding windows, respectively.
Yet, no hardware-aware approach has been proposed that jointly explores the optimal strategy \mvnnok{from a scheduling as well as a hardware point of view}.}

To this end, we present ACCO, \camreadyok{an automated optimizer that explores} \mvnok{efficient} Causal CNN transformation and \lm{DF} scheduling for ST-CNNs on edge 
\mvnnok{hardware} \lm{accelerators}. 
\camreadyok{By cost-modeling the computation and data movement on the accelerator architecture, ACCO automatically selects the best scheduling strategy for the given hardware-algorithm target.}
ACCO\mvnok{'s time-dimension optimization} reaches \mvnnok{a 8.4$\times$ better} Energy-Delay-Product 
\mvnnok{compared to} the fixed \camreadyok{dilated causal conversion}, \mvnok{while ACCO's spatial DF optimization}
\mvnok{improves} $\sim$20\% \mv{compared to} the SotA \mv{DF} exploration toolchain.
When jointly optimizing ST-CNN spatially and temporally, ACCO's \mvnok{scheduling outcomes} are on average 19$\times$ faster and 37$\times$ more energy-efficient than spatial DF schemes.  
\end{abstract}


\copyrightnotice

\section{Introduction} \label{section:introduction}
\begin{figure}[!]
\centerline{\includegraphics[width=\linewidth]{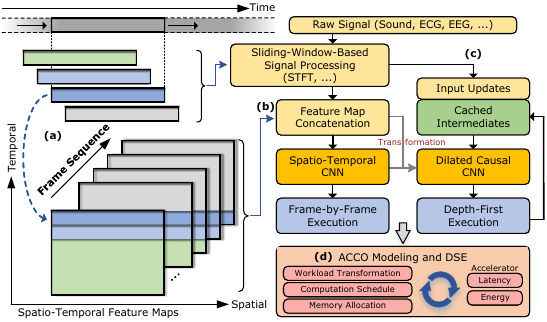}}
\caption{The overview of ACCO: \textbf{a)} \mvnok{illustration} of spatio-temporal input features; \textbf{b)} the conventional execution of ST-CNN; \textbf{c)} the DF execution of dilated causal CNN on input-feature updates; \textbf{d)} \mvnok{ACCO's} accelerator modeling and scheduling design space exploration (DSE). 
}
\label{fig:overview}
\end{figure}

Over the past decade, Convolutional Neural Networks (CNNs) have become an essential workhorse 
for computer vision tasks. 
Carrying forward its success, CNNs have been extended from the image processing field to spatio-temporal reasoning tasks. Recently, many successful models have been developed in different application domains, such as visual tracking \cite{yan2021learning, kumawat2021depthwise}, acoustic perception \cite{zhang2017hello, Kong2019CrosstaskLF}, biomedical information extraction \cite{cho2020spatio, shoeibi2022overview}, etc. 
As shown in Fig. \ref{fig:overview}a, \mvnok{these designs typically} 
include the time axis as an additional dimension of the input feature map 
\mvnok{which then allows to} leverage conventional SotA CNN structures \mvnok{with a reshaped spatio-temporal input}. 
However, adding the required time dimension to these CNNs inflates the CNN models with massive but unnecessary hardware overhead that hinders real-time edge processing. 

Several algorithm-oriented optimization methods are available to ease such hardware pressure, \mv{such as} quantization and pruning for DNN compression, as well as advanced spatio-temporal structures like convolutional recurrent neural networks (CRNN). \mv{Yet, all these optimizations require} additional model re-training steps to avoid accuracy loss.
\mvnok{Spatio-temporal CNN (ST-CNN) can, however, exploit another optimization pass which does not require retraining.} 
\mvnok{Due to (shifted) data repetition across the} 
spatio-temporal input feature maps, \mvnok{there exists} repeated intermediate activations \mvnok{throughout the network across subsequent frames. These} can be cached and reused between adjacent frames, while only the newly updated input lines need to be \mv{propagated through the} 
CNN (Fig. \ref{fig:overview}c).
\mv{This becomes possible by combining principles from} dilated causal CNN topologies \cite{oord2016wavenet} and depth-first (DF) CNN execution \cite{alwani2016fused}. 

\mv{A large amount of computations can be saved from such joint \lm{workload} topology-scheduling optimization \mvnok{without the need for any network retraining. Yet,} this brings new efficiency challenges:}
On the one hand, CNN's parallelism within a single input frame is reduced because the temporal dimension \mvnok{is replaced} by intermediate-result caching. \mv{As a result, the hardware accelerator could be under-utilized, and the weight loading will become more dominant}.
On the other hand, the cached data \mv{puts additional pressure on the memory system, even more so when execution is progressing in batches of several input frames}.

\lm{This results in} \mv{a huge optimization space} 
\camreadyok{\mvnnok{stemming from multiple potential} DNN transformations}
\mv{due to the newly introduced caching and scheduling options.} To tackle this enlarged design space, we propose a unified optimization framework, ACCO, to perform joint \lm{design space exploration (DSE)} on the workload transformation, hardware computation scheduling, and memory allocation (Fig. \ref{fig:overview}d). 
The optimization is based on comprehensive hardware cost modeling \mv{in terms of} energy and latency. 
The contribution\lm{s} of \todo{ACCO} can be highlighted as: 
\begin{enumerate}
    \item \lm{The formal extension of the DF design space} \mv{towards} ST-CNN workloads with an automatic 
    layer-fus\mv{ion} transformation (Section~\ref{section:methodology});
    \item \lm{The \camready{open-source\footnote{\camready{https://github.com/KULeuven-MICAS/ACCO.}}} implementation  of} the DSE on this formalized design space \mv{towards} optimal latency and energy based on precise hardware cost modeling (Section~\ref{section:implementation});
    \item \lm{The conduction of four} case studies, including ablation studies and SotA comparisons to demonstrate ACCO's strength, \lm{optimizing} representative ST-CNN models for both the single and batch \lm{frame(s)}
    (Section~\ref{section:experiments}).
\end{enumerate}


\section{Background and Motivation} \label{section:background}

\subsection{Dilated Causal CNNs} \label{section:background_algorithm}

\mvnok{Temporal data can be efficiently processed through} dilated causal convolutions \cite{oord2016wavenet}. 
As shown in Fig. \ref{fig:background}a, these extend the receptive field of a \lm{CNN} \mvnok{through} dilation and causality, without increasing the number of parameters. 
By introducing gaps in the filters, the dilation factor controls the spacing between the filter elements, enabling larger receptive fields. 
Causality ensures that information flows only from the past to the future, making it suitable for sequential data processing. This combination allows the model to capture long-range dependencies efficiently, making dilated causal convolution popular in tasks such as natural language processing, audio processing, and time series analysis.
For example, the TCN \cite{lea2016temporal} structure leverages such structure to efficiently process sequential data, such as time series or temporal sequences.

\subsection{Depth-First DNN Execution} \label{section:background_depthfirst}
\begin{figure}[!]
\centerline{\includegraphics[width=0.95\linewidth]{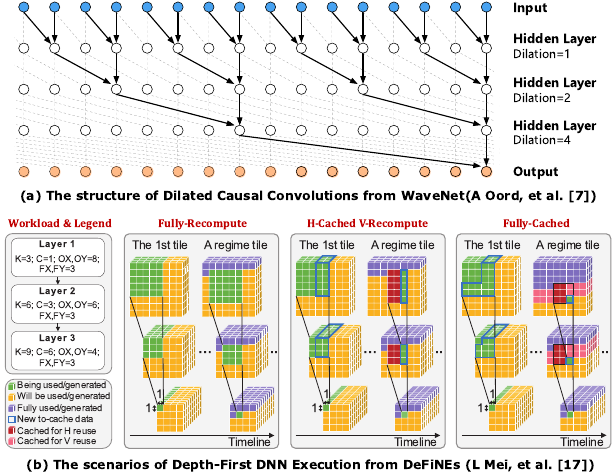}}
\caption{The background concept diagrams: (a) the dilated causal convolution structure \cite{oord2016wavenet}; (B) the Depth-First DNN execution scheduling \cite{mei2023defines}. 
}
\label{fig:background}
\end{figure}

\mvnok{Spatial CNNs can exploit} \lm{DF execution (a.k.a layer fusion \cite{alwani2016fused}). In contrast to layer-by-layer execution, \mvnok{this technique} breaks each layer into multiple \mvnok{individually schedulable} tiles and schedules them across layers, \mvnok{while either caching or recomputing data required by multiple tiles} (Fig. \ref{fig:background}b).}
DF scheduling can \todo{largely benefit in terms of}
energy efficiency and latency 
\todo{because of the \mvnok{its reduced memory footprint stemming from} in-place consumption of intermediate DNN activations between layers.}
\todo{\mvnok{Since the smaller intermediate activation tiles} 
can be stored and reused at a lower memory level (e.g., the global buffer) 
\todo{\mvnok{they save a lot of costly memory fetches} from
higher memory levels (e.g., DRAM).}
}

\lm{DF execution contains various scheduling options and leads to a large design space\todo{, such as} different tile sizes, number of fused layers, cross-tile overlapped data handling, etc.} 
\todo{Different DF schemes can be observed in many DNN accelerators published recently} 
\cite{15FusedLayerCNN_2016MICRO,18AFullHD60_2019VLSI,23A121T_2022JSSCC,DepFiN_2021VLSI,16ECNN_2019MICRO}. 
\lm{
\todo{Yet, all} these implementations \todo{have ad-hoc and} pre-defined tile sizes, fusing depth, and overlap data storing modes. It is challenging to justify whether a design choice is optimal for target workloads \todo{due to the immense DF design space}, necessitating automatic and fast DSE frameworks.}

\subsection{Automatic Scheduling Exploration Frameworks} \label{section:background_frameworks}

\lm{In exploring the DNN scheduling space automatically and quickly, two key \todo{\mvnok{aspects} are highlighted: evaluation and exploration.}
The former \mvnok{aspect} models different scheduling options
\todo{while rapidly evaluating their} hardware costs. 
The latter \mvnok{aspect} explores the DNN scheduling space and automatically generates
\todo{candidate} schedules
for cost evaluation. This search-evaluation loop can iterate multiple times to converge towards the Pareto-optimal scheduling points.}

\lm{Several DF scheduling exploration frameworks have been developed recently, 
such as~\cite{DNNFuser,xing2019dnnvm,mei2023defines,MCUNetV2},
each with a different focus.}
\lm{For example, DNNFuser~\cite{DNNFuser} focuses on the DF space search and proposes a one-shot Transformer-based DF scheduler;
DNNVM~\cite{xing2019dnnvm} transforms DNN models into DAG and enumerates all potentially profitable fusion opportunities by a heuristic subgraph isomorphism algorithm;
MCUNetV2~\cite{MCUNetV2} introduces patch-based inference and jointly optimizes the neural architecture and DF scheduling so as to reduce the peak memory usage;
DeFiNES~\cite{mei2023defines} constructs a 3-axis DF scheduling design space and proposes a unified analytical modeling approach for fast hardware cost evaluation.}

\camreadyok{Yet, such automatic exploration remains unaware of the causal input relationship as mentioned in Section \ref{section:background_algorithm}.}

\subsection{\lm{Motivation}} \label{section:motivation}

\todo{Fig. \ref{fig:background} \mvnok{shows the similarity in dataflow}} 
between the dilated causal convolution structure and the DF execution scheduling. 
\camreadyok{This sheds light on efficiently optimizing ST-CNN workloads in a common framework, as indicated in} 
Fig. \ref{fig:concept_relationship}c and \ref{fig:concept_relationship}d, 
for further performance/efficiency \lm{boosting}. 

\camreadyok{To do so, new challenges can be identified as follows:}
\begin{enumerate}
    \item to convert the \camreadyok{causal} overlap \mvnok{across the time-dimension} into a DF schedulable workload while maintaining mathematical equivalence with the original ST-CNN;
    \item to handle the difference in cost modeling and optimization between a single causal input frame and the entire flattened spatio-temporal batch \lm{of frames};
    \item to jointly manage the increased size of the intermediate activation cache as well as the weights under the limited on-chip memory resources.
\end{enumerate}

ACCO \lm{tackles these challenges and \mvnok{enables to automatically} explore the enlarged scheduling space (Section \ref{section:methodology}).}

\section{Design of ACCO} \label{section:methodology}

This section \mv{details the ACCO concepts \todo{and components} for} the scheduling optimization of causal CNNs. 
Firstly, Section~\ref{section:methodology_concept} introduces the relationship between \mv{the core concepts on which ACCO is based,} such as the ST-CNN, dilated causal CNN structure, and DF scheduling.
Next, Section~\ref{section:methodology_transformation} proposes a general methodology to transform \mv{a} ST-CNN into a dilated causal CNN structure \mv{compatible with} DF scheduling.
Section~\ref{section:methodology_design_space} \mv{summarizes} the design parameters that form ACCO's design space, while Section~\ref{section:methodology_tradeoff} identifies the corresponding trade-offs.

In the following sections,
we focus on 2D ST-CNNs with a row-major execution order. 
\mv{The concepts are \lm{also} expandable towards higher dimensionality.}
As shown in Fig. \ref{fig:concept_relationship}a and \ref{fig:concept_relationship}c, the feature map's 
$\mathbf{X}$ dimension is for the spatial, and $\mathbf{Y}$ dimension is for the temporal \mvnok{dimension}. 

\subsection{\mv{Core} Concepts of ACCO} \label{section:methodology_concept}

\begin{figure}[!]
\centerline{\includegraphics[width=\linewidth]{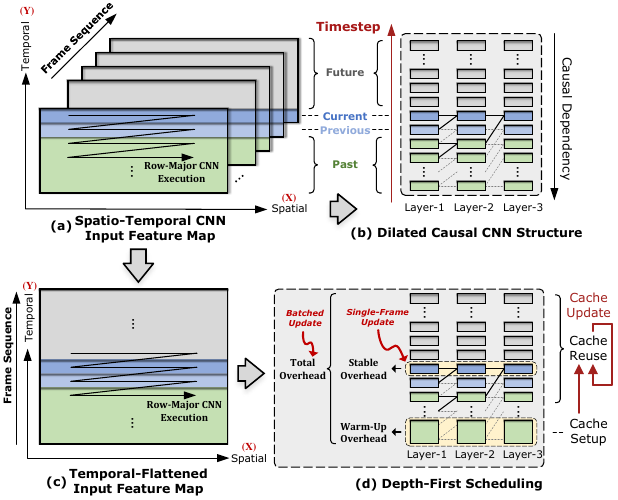}}
\caption{ACCO's basic concept relationship: \textbf{a)} the original input feature map for ST-CNNs; \textbf{b)} the dilated causal CNN transformation that reduces temporally-overlapped computation; \textbf{c)} \camready{the input flattening that embeds such temporal-overlapping information for scheduling optimization;}
\textbf{d)} the hardware-oriented DF scheduling that optimizes the \camready{flattened ST-CNN}.
}
\label{fig:concept_relationship}
\end{figure}

As introduced in Section~\ref{section:introduction}, conventional ST-CNNs treat the temporal feature dimension in the same way as the spatial dimension. 
However, \mv{as visualized} in Fig. \ref{fig:concept_relationship}a, the \mv{overlap between values in the} feature map of \mvnok{subsequent} input frames indicates the \mv{potential} reuse of previously computed CNN activations. 

\camready{As introduced} in Section~\ref{section:background_algorithm}, a dilated causal CNN structure can exploit such intermediate result caching and reuse it across sequential ST-CNN input frames.
\mv{This is illustrated in} Fig. \ref{fig:concept_relationship}b. Only the latest row of the current input is computed across the CNN. In contrast, the remainder of the input feature map is already calculated in previous time steps and thus reused according to the causal dependency of different layers.
Chronologically, such execution can be seen as the continuous processing of the CNN on a temporal-flattened input feature map, as shown in Fig. \ref{fig:concept_relationship}c.

\camready{With such causal overlapping  embedded in the flattened feature map,}
DF execution \mv{along} the temporal feature dimension (Fig. \ref{fig:concept_relationship}d) \mv{can be exploited}.
Both the DF and dilated causal CNN are designed to minimize the size and lifetime of every layer's input data by serializing the computation and reusing shared intermediate results. That is, each input feature or intermediate activation only persists in the system until all the dependent computations have been completed for all outputs. 
\camready{Unlike the algorithm-oriented dilated causal CNN, the DF scheduling considers 
both the hardware cache prefill phase (``Warm-Up'') and the cache update phase (``Stable'').}

\camready{
For the real-time single-frame update, only one temporal slice has to be computed (Fig. \ref{fig:concept_relationship}d).
Likewise, the batched update is for trigger-based processing on collected snippets (e.g., sound events~\cite{Kong2019CrosstaskLF} and biomedical sequences~\cite{cho2020spatio, shoeibi2022overview}).
}

\subsection{ST-CNN Transformation} \label{section:methodology_transformation}
Yet, DF execution \mv{so far has only been developed for frame-based processing, lacking} the ability to track temporal relationships \mvnok{across different sequential input frames of the} original ST-CNN (Fig. \ref{fig:concept_relationship}a).
Therefore, \mv{ACCO introduces a} workload transformation \mv{step} 
to enable the automated scheduling optimization exploiting both temporal computational reuse, as well as DF execution. ACCO first performs a transformation on the original ST-CNN \mv{to obtain, on the one hand, a temporally-flattened input feature map (Fig. \ref{fig:concept_relationship}c) and, on the other hand, an equivalent causal CNN operating on that input map} 
\camreadyok{for DF scheduling (Fig. \ref{fig:concept_relationship}d).}
\mv{This transformation is possible for any network consisting of} sliding-window-based kernels (CONV, POOL, Residual, etc.). 
\mv{Other} 
kernels, such as the Fully-Connected layers, are \mv{left for future research.}

\camready{For the transformed input, we denote its temporal stride as $\mathcal{S}_0$, which links the temporally-flattened input (Fig.~\ref{fig:concept_relationship}c) back to its original form (Fig.~\ref{fig:concept_relationship}a). That is, with each new original input, the flattened input map grows with $\mathcal{S}_0$ in size at the temporal dimension. 
}

\begin{figure}[!]
\centerline{\includegraphics[width=\linewidth]{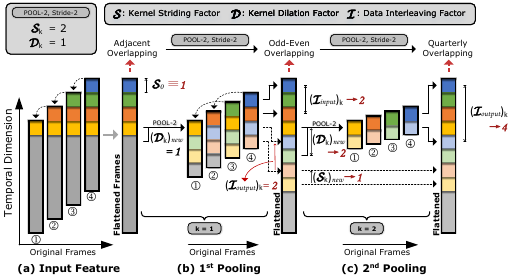}}
\caption{\camready{A 2-pooling example on how ACCO fixes the layer-stride mismatch at the ST-CNN Transformation by converting strides to dilations.
The diagram explains the processing of 4 consecutive ST-CNN input frames in the original and flattened manner.
Such conversion is meant for result equivalence after the frame flattening.
The colorized data alignment highlights unique values.
}
}
\label{fig:causal_cnn_transformation}
\end{figure}

\mv{Further, for the \camready{ST-CNN} model transformation, special care has to be given to layers \camready{whose temporal stride size} $\mathcal{S}_k$ 
mismatches with the \camready{global input's $\mathcal{S}_0$.}
}
Fig. \ref{fig:causal_cnn_transformation} provides an \camready{$\mathcal{S}_0=1$ example for such temporal stride mismatch and how it is resolved after transformation.} 
If \mv{any intermediate} layer has $stride \neq \mathcal{S}_0$ in the same dimension,
\camready{
the overlapping pattern of input data would be changed at the output. 
For example, after a Stride-2 POOL-2 layer, the overlapping frames \circled{1}{0.7} and \circled{2}{0.7} in Fig.~\ref{fig:causal_cnn_transformation}a become mutually exclusive in Fig.~\ref{fig:causal_cnn_transformation}b, while frame \circled{1}{0.7} and \circled{3}{0.7} keeps their overlapping. 
In our flattened model, we first align all these data in the chronological order of sliding-window computation to ensure no data is missing during the DF scheduling. 
Then, we perform the stride-to-dilation transformation to help the next layers correctly fetch their corresponding input from these flattened feature maps.
}

\mv{More formally,} considering a Layer-$k$ in an $N$-layer ST-CNN model, we denote its original temporal-dimension stride and dilation factor as $\boldsymbol{\mathcal{S}}_k$ and $\boldsymbol{\mathcal{D}}_k$, respectively.
The original input frame updates on the temporal dimension with stride size  $\boldsymbol{\mathcal{S}}_0$.
\mv{Therefore, the equivalent} Layer-$k$ \mv{of the flattened model \mvnok{after transformation} is} defined as:
\begin{equation} \label{eq:stride_to_dilation_general}
    \left\{
    \begin{aligned}
        (\boldsymbol{\mathcal{S}}_k)_{new}&=\gcd(\boldsymbol{\mathcal{S}}_0, \boldsymbol{\mathcal{S}}_k) & \\
        (\boldsymbol{\mathcal{D}}_k)_{new}&=\frac{(\boldsymbol{\mathcal{I}}_{input})_k}{(\boldsymbol{\mathcal{S}}_k)_{new}} \times \boldsymbol{\mathcal{D}}_k & ,k \in [1, N] \\
        (\boldsymbol{\mathcal{I}}_{output})_k&=\frac{(\boldsymbol{\mathcal{I}}_{input})_k}{(\boldsymbol{\mathcal{S}}_k)_{new}} \times \boldsymbol{\mathcal{S}}_k
    \end{aligned}
    \right.
\end{equation}
where the $\boldsymbol{\mathcal{I}}_{input}$ and $\boldsymbol{\mathcal{I}}_{output}$ denote the amount of interleaved original frames in the transformed Layer-$k$'s flattened input and output, respectively.  
As illustrated in Fig. \ref{fig:causal_cnn_transformation}b and \ref{fig:causal_cnn_transformation}c, this factor is passed down into the successor layer (e.g., Layer-($k$+1)) as $(\boldsymbol{\mathcal{I}}_{input})_{k+1}=(\boldsymbol{\mathcal{I}}_{output})_{k}$. 

In practice, we normalize to $\boldsymbol{\mathcal{S}}_0 \equiv 1, \boldsymbol{\mathcal{I}}_0 \equiv 1$ as the global input's generation rate.
Therefore, according to Eq.\eqref{eq:stride_to_dilation_general}, the new stride and dilation factors can be computed: 
\begin{equation} \label{eq:stride_to_dilation_actual}
    \left\{
    \begin{aligned}
        (\boldsymbol{\mathcal{S}}_k)_{new} &\equiv 1 & \\
        (\boldsymbol{\mathcal{D}}_k)_{new}&= \boldsymbol{\mathcal{D}}_k \times \prod _i \boldsymbol{\mathcal{S}}_i & ,i \in \mathbf{P}  \\
    \end{aligned}
    \right.
\end{equation}
with $\mathbf{P}$ the index set of all the predecessor layers to Layer-$k$. 


\camready{
With this transformation, temporal features of the ST-CNN computation are fully embedded into one spatially-defined model and 
can be exploited during DF scheduling.
}

\subsection{DF Scheduling Design Space} \label{section:methodology_design_space}



Compared to the original frame-by-frame CNN execution, ACCO aims to find Pareto-optimal DF scheduling strategies for the transformed ST-CNN workload.
Given a hardware accelerator architecture with a fixed spatial-unrolling and limited on-chip memory resources, ACCO explores the design space with two key \mv{exploration} parameters: 

\subsubsection{\textbf{Computation Tiling Size}}
CNN kernels can be described as nested for-loops across all \mv{kernel and output tensor} dimensions. 
\mv{Splitting any of these for-loops into an inner for-loop and an outer for-loop and reordering them allows for workload tiling to optimize memory usage.} 
ACCO performs a tiling exploration for each layer's output dimensions, \mv{exploring loop splitting with configurable} \textbf{Spatial Tiling Size} and the \textbf{Temporal Tiling Size} \mv{along the spatial and temporal dimensions of} the temporal-flattened feature map of Fig. \ref{fig:concept_relationship}c. The tiling along these two dimensions directly \mv{influences} the DF \mvnok{scheduling granularity, impacting the} computing and caching overhead, as introduced in Section \ref{section:background_depthfirst}. 
DF scheduling partitions the last layer's output features based on these tiling sizes and \mvnok{automatically} collects dependent computation from previous layers.

\subsubsection{\textbf{Depth-First Layer Fusion Range}}
A DNN workload can be cut and fused \mv{in multiple ways into a set} of sub-stacks on which DF scheduling is performed separately. The \mv{location of these stack cuts}, i.e., the \textbf{Layer Fusion Range}, \mv{is the second exploration parameter of ACCO}. 

\subsection{Design Space Trade-off Identification} \label{section:methodology_tradeoff}


\begin{figure}[!]
\centerline{\includegraphics[width=\linewidth]{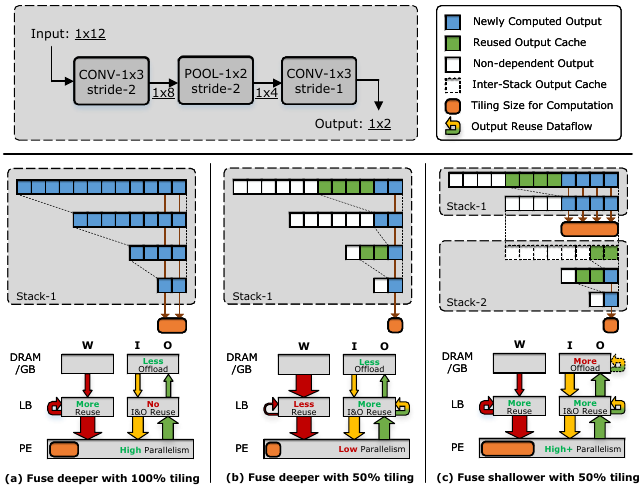}}
\caption{\camready{A 3-layer workload example on the ACCO scheduling trade-offs. The tile-wise computation of three manually-chosen corner cases is illustrated: \textbf{a)} DF fusion of all layers, 100\% tiling at output; \textbf{b)} DF fusion of all layers, 50\% tiling at output; \textbf{c)} DF fusion into 2 sub-stacks, 50\% tiling at each sub-stack's output. 
Their hardware overheads are qualitatively visualized assuming the ideal data localization.}
Abbreviations: \textbf{W}/\textbf{I}/\textbf{O} = Weights/Inputs/Outputs; \textbf{GB} = Global Buffer; \textbf{LB} = Local Buffer; \textbf{PE} = Processing Elements.
}
\label{fig:scheduling_design_space}
\end{figure}

Based on the introduced exploration parameters, ACCO explores the Pareto-optimal solutions for any \lm{CNN} workload-accelerator pair.

Intuitively, one should tile as much as possible and fuse layers as deep as possible 
to maximize the computational parallelism and the reuse of cached data, just like the \mv{traditional} dilated causal CNN of Fig. \ref{fig:background}a. 
However, the actual situation is much more complex.

To \mv{illustrate} this entangled design space, Fig. \ref{fig:scheduling_design_space} shows an example with 1d input data and a 3-layer dummy CNN workload (Conv3-s2, Pool2-s2, and Conv3-s1). The input and output sizes are set to 1$\times$12 and 1$\times$2, respectively. Three \camready{manually-chosen} corner cases with \camready{typical} tiling and layer fusion \mvnok{are visualized} \camready{on their tile-wise computation complexity and qualitative hardware overhead}.

To begin with, Fig. \ref{fig:scheduling_design_space}a represents the aforementioned \mv{baseline} case, where all layers are fused into one DF stack \mv{without} tiling the output \mv{(tile size = output size)}.
\mv{In this case,} 
the scheduling falls back into normal layer-by-layer \mv{frame-by-frame}  execution, where no 
intermediate result is cached. 
\mv{Of course, the resulting parallelism within each layer allows the weights to be} 
reused or consumed locally. 

Fig. \ref{fig:scheduling_design_space}b opts for a smaller output tile size under the same end-to-end layer fusion to \mv{exploit} DF \mv{processing benefits}. 
\mv{Intermediate feature maps are cached and consumed as soon as possible, leading to more efficient IO memory usage}. 
\mv{This comes at the expense of reduced intra-layer parallelism}, leading to poor weight data reuse and low PE utilization.

Finally, Fig. \ref{fig:scheduling_design_space}c \mvnok{pursues} \mv{tiled DF execution with} shallower layer fusion that cuts the workload into two \mv{separate} sub-stacks.
In the Stack-1 with wider output size, a higher tiling size can be chosen while still preserving DF features. 
\camready{
Compared to Fig. \ref{fig:scheduling_design_space}b, the active data cache (green) in the IO memory stays the same because of the same 50\% DF tiling ratio, while 
more computation parallelism (orange) and weight reuse (red) \lm{are} \mv{obtained}.
\mv{Of course, this solution comes with additional} data movement to transfer cached data between these sub-stacks. 
Therefore, ACCO performs a DRAM Skipping (\textbf{DS}) check to improve inter-sub-stack data locality.
}

In short, the actual Pareto-optimal strategy might vary for different workloads and accelerator architectures. \mv{A careful exploration with a reliable cost model is required to explore the complete} design space. \mv{This will be realized in the ACCO implementation} (Section IV).

\begin{figure}[!]
\centerline{\includegraphics[width=0.75\linewidth]{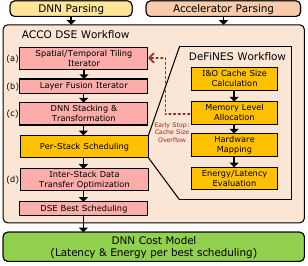}}
\caption{The overview of ACCO's implementation. 
}
\label{fig:dse_overview}
\end{figure}

\begin{table}[t]
    \centering
    \caption{Workload parameter relationship between the three ST-CNN execution modes. 
    }
    \label{tab:workload_parameter_per_mode}
    \begin{threeparttable}
    \renewcommand\arraystretch{1.3}
    \begin{tabular}{r|c|c|c}
    \toprule[0.22ex]
        \textbf{Execution Mode}& \textbf{Baseline}& \textbf{Real-Time}\tnote{2} & \textbf{Batch}\tnote{2}\\ \Xhline{0.22ex}
         \textbf{Spatial Output} & $OX\tnote{1}$ & $OX$ & $OX$\\ \hline
         \textbf{Temporal Output} & $OY\tnote{1}$ & 1 & $OY \times \prod \boldsymbol{\mathcal{S}}_i$ \\ \hline
        \textbf{\camready{Temporal Frames}} & \camready{$\prod \boldsymbol{\mathcal{S}}_i$}& $\camready{1}$ & $\camready{1}$\\
    \bottomrule[0.22ex]
    \end{tabular}
    \begin{tablenotes}
        \item[1] The original ST-CNN's output size. X/Y for spatial/temporal features as defined in the prologue of Section \ref{section:methodology}.
        \item[2] The real-time/batch mode which represents the workload of \camready{single-frame/batched  update 
        in Fig. \ref{fig:concept_relationship}d. Hence, one \textbf{Batch}-mode temporal frame is equivalent to \camready{$\prod \boldsymbol{\mathcal{S}}_i$} \textbf{Baseline}-mode frames.}
    \end{tablenotes}
    \end{threeparttable}
\end{table}

\begin{table}[t]
\centering
\setlength{\abovecaptionskip}{0.cm}
\caption{The parameters of the evaluated DNN workloads.}
\vspace{-0.cm}
\label{tab:dnn_workload}
\includegraphics[width=\linewidth]{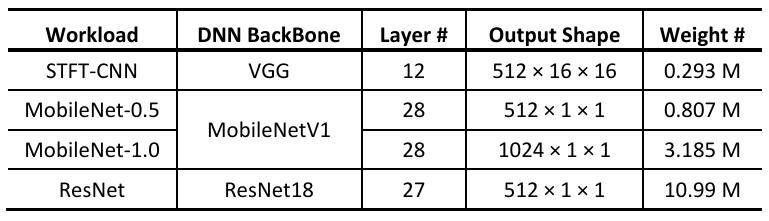}
\end{table}

\section{Implementation of ACCO} \label{section:implementation}

To enable automatic workload transformation and scheduling exploration on the formalized design space from Section~\ref{section:methodology}, we build ACCO \mvnok{as an extension to} DeFiNES \cite{mei2023defines}. DeFiNES is a silicon-verified scheduling exploration tool that enables the rapid DF optimization on DNN workloads given the accelerator PE architecture and on-chip memory hierarchy.

The overview of ACCO's DSE workflow is shown in Fig.~\ref{fig:dse_overview}.
The system is built with new stages on top of DeFiNES to implement Section~\ref{section:methodology}'s concept accordingly: a) The tiling exploration, b) the layer fusion exploration, c) the ST-CNN transformation, d) the inter-sub-stack data transfer optimization. 
For each fused DF sub-stack, the DeFiNES toolchain is invoked for intra-sub-stack DSE. 
To speed up the enlarged DSE process, an early-stopping module is added to prevent \mvnok{exploring schedules} \mv{with activation tiles that do not fit into on-chip memory}. 
Finally, all evaluated layer sub-stacks are collected and \mv{aggregated} for end-to-end cost modeling.
In Section~\ref{section:experiments}, ablation studies are carried out to show the efficacy of each module.

\section{Experiments and Discussions} \label{section:experiments}
We carry out four case studies in this section to verify the effectiveness of ACCO. Section \ref{section:experiments_setup} introduces our experiment setups. 
Section \ref{section:experiments_case_study_1} and \ref{section:experiments_case_study_2} conducts two ablation studies on ACCO's DSE tradeoffs and overall results. 
Section \ref{section:experiments_case_study_3} compares ACCO with its SotA baseline DeFiNES \cite{mei2023defines}. 
Section \ref{section:experiments_case_study_4} demonstrates ACCO's benefits for ST-CNNs in real-time scenarios against other SotA solutions.

\begin{table*}[t]
\centering
\setlength{\abovecaptionskip}{0.cm}
\caption{The parameters of the evaluated hardware accelerator architectures.}
\vspace{-0.cm}
\label{tab:hardware_architecture}
\includegraphics[width=0.8\linewidth]{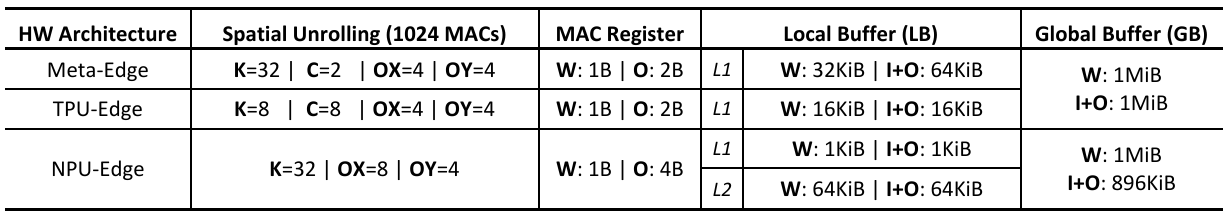}
\end{table*}

\begin{figure*}[!]
\centerline{\includegraphics[width=\linewidth]{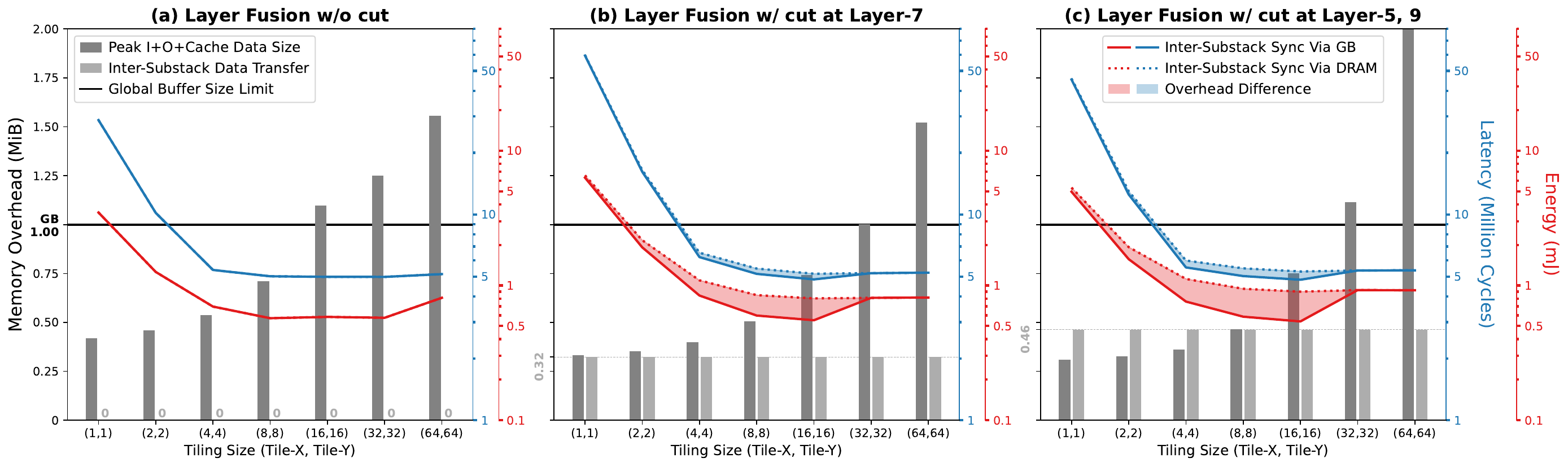}}
\caption{The ablation study of ACCO’s design space trade-offs across 7 \textbf{spatio-temporal tiling cases} and 3 \textbf{layer-fusion scenarios} (w/o cut, cut at Layer-7, and cut at Layer-5,9), with \textbf{batch-mode} STFT-CNN workload on the Meta-Edge accelerator. Two aspects are studied: 1) how the design parameters of tiling and layer fusion impact the data size of Peak I+O+Cache and Inter-Substack transfer; 
2) how the overall latency-energy trade-off reacts to these design parameters. 
Abbreviation: \textbf{GB} = Global Buffer. 
}
\label{fig:stftCNN-16x-ablation}
\end{figure*}

\subsection{Experiment Setups} \label{section:experiments_setup}
We evaluate three execution modes for ST-CNN workloads: \textbf{Baseline}, \textbf{Real-time}, and \textbf{Batch}, computing on the original input, single time-step input update, and the temporal-flattened input, respectively. 
Following to the transformation discussed in Section~\ref{section:methodology_transformation}, the workload parameters are listed in Table~\ref{tab:workload_parameter_per_mode}. 
Note that such \textbf{Real-time} computation is mathematically equivalent to the \textbf{Baseline} single-frame execution.

We define four workloads for \mv{our} case studies in Table \ref{tab:dnn_workload}. These workloads leverage the following backbone structures: VGG \cite{simonyan2014very}, MobileNetV1 \cite{howard2017mobilenets}, and ResNet18 \cite{he2016deep}, which are widely adopted for feature extraction in spatio-temporal applications \cite{cho2020spatio, shoeibi2022overview, zhang2017hello, Kong2019CrosstaskLF, yan2021learning, kumawat2021depthwise}. The data bit width is set to 8~\mv{bits in accordance with} quantized edge CNNs.
Each model excludes its \mvnok{final} fully-connected layers, as explained in Section \ref{section:methodology_transformation}. 

As the DSE hardware target, we select three well-known DNN accelerator architectures: Meta\mv{'s architecture} \cite{sumbul2022system}, the Edge TPU \cite{yazdanbakhsh2021evaluation}, and the Tesla NPU \cite{talpes2020compute}. 
\mv{To mimic} edge execution context and \mv{ensure a} fair comparison, we normalize their compute array dimension to 1024 \mvnok{MAC} units with an on-chip global buffer capacity of 2 MiB, while maintaining each accelerator's spatial unrolling strategy and memory hierarchy. The \mv{resulting} architectures \mv{are summarized} in Table \ref{tab:hardware_architecture}.
All hardware modeling follows the setup in the baseline toolchain \cite{mei2023defines} with CACTI-based \cite{balasubramonian2017cacti} \mv{memory models}.

For convenience, we list common term abbreviations used \mv{throughout} this section \mvnok{for result discussion} as follows:
\begin{enumerate}
    \item \textbf{LBL}: the basic Layer-by-Layer DNN execution;
    \item \textbf{DF}: the Depth-First DNN scheduling \cite{alwani2016fused} 
    \item \textbf{FT}: the Fixed-Tiling strategy that \camreadyok{tiles each DF computation by HW architecture's unrolling factors (Table \ref{tab:hardware_architecture});} 
    \item \textbf{TE}: ACCO's spatial/temporal Tiling Exploration;
    \item \textbf{LFE}: ACCO's DF Layer Fusion Exploration;
    \item \textbf{DS}: ACCO's inter-sub-stack DRAM Skipping strategy; 
    \item \textbf{Best}-\textbf{LAT/ENE/EDP}: DSE \camreadyok{cost functions} that optimize for the best Latency, Energy or Energy-Delay-Product;
\end{enumerate}


\subsection{Case Study 1: The Ablation Study of ACCO DSE Tradeoffs} \label{section:experiments_case_study_1}

To quantify the trade-off discussion of Fig. \ref{fig:scheduling_design_space}, we design this ablation study to manifest the trade-offs of each ACCO DSE component by optimizing the batch-mode STFT-CNN workload on Meta-Edge hardware.

\mvnok{The exploration results across ACCO's degree of freedom (tiling size and DF stack depth)} 
are shown in Fig. \ref{fig:stftCNN-16x-ablation}. The results can be \mvnok{interpreted} as:

\subsubsection{\textbf{Tiling Exploration (TE)}}

Generally, all 3 scenarios in Fig. \ref{fig:stftCNN-16x-ablation} prove the benefits of workload parallelization from \mvnok{increased tiling sizes}.  
However, \mvnok{larger tiles} also increase the input and output data volume of each tile, which is further magnified by the caching requirement in DF execution (i.e. the Peak I+O+Cache in Fig. \ref{fig:stftCNN-16x-ablation}).

For the STFT-CNN, the tiling benefit saturates around the tile size (16,16) after achieving about 8$\times$ latency and 6$\times$ energy savings on average. This testifies to the case in Fig. \ref{fig:scheduling_design_space}a because \camreadyok{when the tiling \mvnnok{increases beyond the} STFT-CNN's output shape (16$\times$16), the entire layer is \mvnnok{processed in one go,} thus the DF computation falls back to be layer-by-layer.}
Especially, the energy overhead of Tiling-(64,64) in Fig. \ref{fig:stftCNN-16x-ablation}a worsens by 40\% because \camreadyok{DF still tries to manage all layers' data movement as a layer-fused block so that the LB overflows at this corner.}

\subsubsection{\textbf{Layer-Fusion Exploration (LFE)}}

The direct benefit of cutting the entire CNN into multiple sub-stacks is \mvnok{the reduction in} active data volume.
Comparing Fig. \ref{fig:scheduling_design_space}c to \ref{fig:scheduling_design_space}a, the data engaged by the tile-unit computation reduces after cutting the workload into 2 sub-stacks (more non-dependent data).

Comparing the Peak I+O+Cache overhead of Fig. \ref{fig:stftCNN-16x-ablation}b\&\ref{fig:stftCNN-16x-ablation}c to \ref{fig:stftCNN-16x-ablation}a, an average reduction of 20.5\% can be observed.
However, the latency-energy benefits from \textbf{LFE} would be neutralized by the additional inter-sub-stack data transfer via DRAM. 

\subsubsection{\textbf{Inter-sub-stack DRAM Skipping (DS)}}

ACCO's baseline design, DeFiNES \cite{mei2023defines}, synchronizes inter-sub-stack intermediate results through DRAM. Yet, from the analysis in \textbf{2)}, the Peak I+O+Cache of the entire workload can be within the size of Global Buffer (1 MiB) if performing the \textbf{TE} and \textbf{LFE} wisely (scenarios in Fig. \ref{fig:stftCNN-16x-ablation}b\&\ref{fig:stftCNN-16x-ablation}c with below-(32,32) tiling).

That is, in these cases, the intermediate activations could reside in GB to be used by the next sub-stack, skipping the DRAM detour. 
Hence, the additional latency and energy overhead of such data transfer can be saved, marked by the \mvnok{red and blue} shades in Fig. \ref{fig:stftCNN-16x-ablation}b\&\ref{fig:stftCNN-16x-ablation}c, \mvnok{with varying} benefits \mvnok{across} different choices in \textbf{TE} and \textbf{LFE}. 

\begin{figure}[!]
\centerline{\includegraphics[width=.9\linewidth]{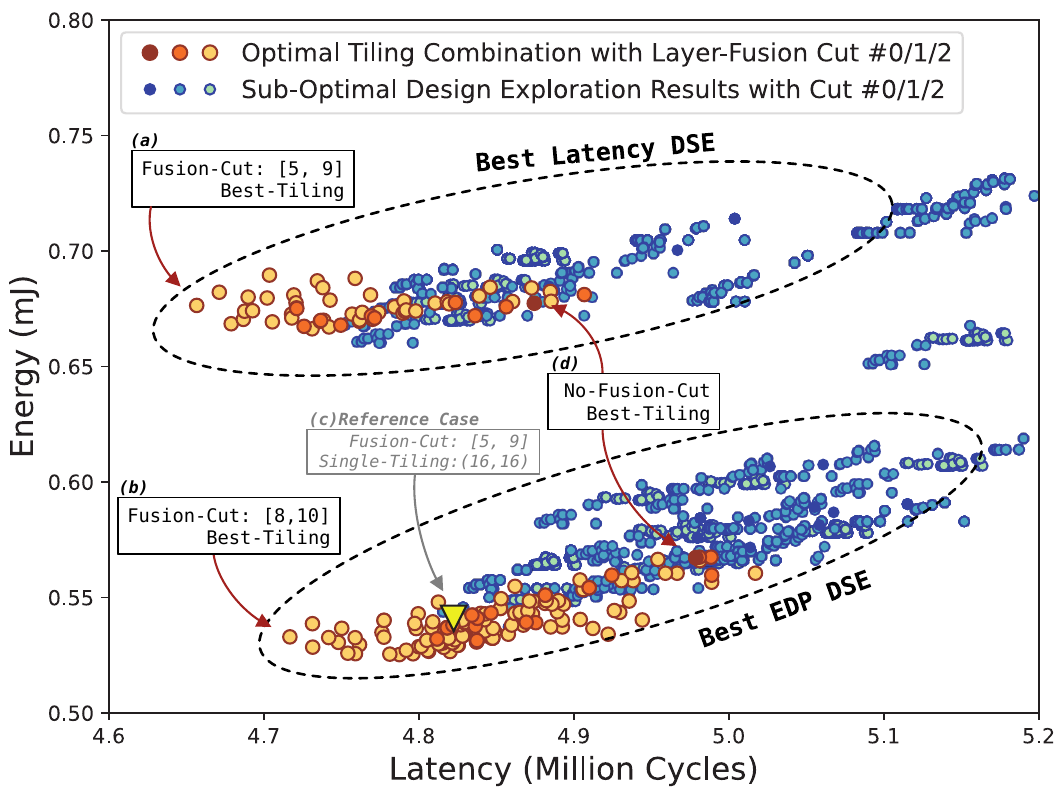}}
\caption{The ACCO DSE result ensemble for \textbf{batch-mode} STFT-CNN workload on the Meta-Edge accelerator. The maximal Layer-Fusion sub-stack amount explored is 3 (Cut \#0/1/2). Under each layer-fusion scenario, the best combination of tiling strategy for each sub-stack is explored towards the DSE target metrics (e.g., Latency and Energy-Delay-Product (\textbf{EDP})).
The best trade-off from Fig. \ref{fig:stftCNN-16x-ablation} is shown as the \camready{sub-optimal reference case (c)}.
}
\label{fig:stftCNN-16x-DSE}
\end{figure}

\subsection{Case Study 2: The Best Outcome from Joint DSE} \label{section:experiments_case_study_2}

\begin{figure}[!]
\centerline{\includegraphics[width=\linewidth]{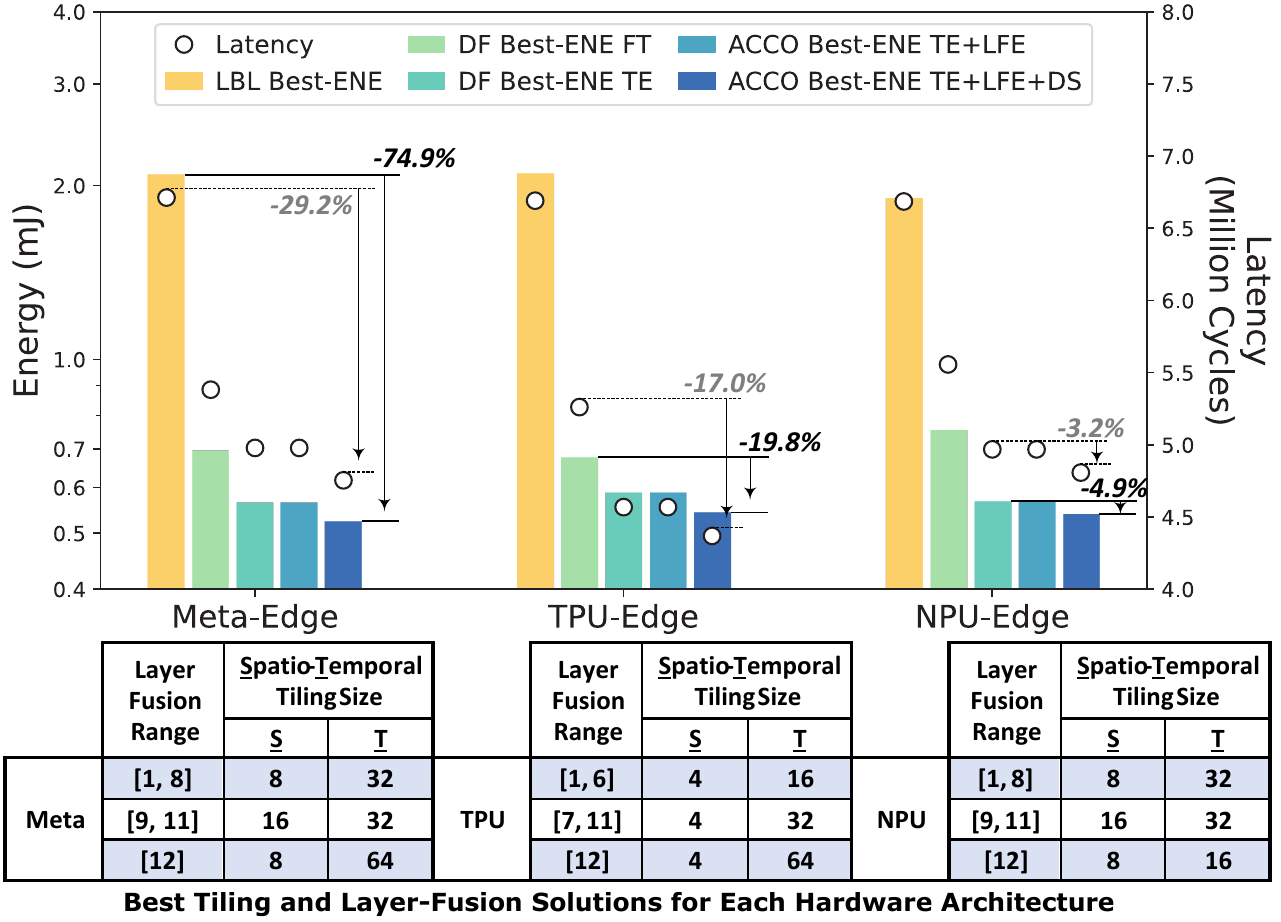}}
\caption{The energy-latency ablation study of \textbf{batch-mode} STFT-CNN DSE on all three accelerators, together with the best scheduling solutions found by ACCO. Compared with the ACCO scheduling, the Layer-By-Layer (\textbf{LBL}) and basic DF (\textbf{DF}) execution patterns are shown as the baseline. 
}
\label{fig:stftCNN-16x-best}
\end{figure}

Fig. \ref{fig:stftCNN-16x-DSE} \mvnok{plots a range of ACCO exploration results for best latency and best energy explorations in batch mode, highlighting also the best results analyzed in} Fig. \ref{fig:stftCNN-16x-ablation}. 
\mvnok{It is clear that more optimal solutions can be found with} ACCO's \mv{complete} DSE strategy (\textbf{TE}+\textbf{LFE}+\textbf{DS}). 
The Pareto-optimal frontier can be pushed forward by 3.3\% for Best-LAT (Fig. \ref{fig:stftCNN-16x-DSE}a) and 4.1\% for Best-EDP (Fig. \ref{fig:stftCNN-16x-DSE}b) compared to the reference case (Fig. \ref{fig:stftCNN-16x-DSE}c).

Next, we perform the joint DSE for batch-mode STFT-CNN on all the modeled hardware architectures.
As the baseline comparison and proof of concept, we also evaluate Layer-by-Layer execution and basic DF optimization on the same temporal-flattened workload. \mvnok{The results are summarized in} Fig. \ref{fig:stftCNN-16x-best}, where the best overall scheduling strategy given by ACCO, \mvnok{shows to be} on average, 14.3\% faster and 24.2\% more energy-efficient compared to the DF baseline.
Compared with the LBL baseline, this benefit further increases to 30.6\% for latency and 73.3\% for energy.

\subsection{Case Study 3: SotA Comparison with Baseline Toolchain} \label{section:experiments_case_study_3}

As introduced in Section \ref{section:background_frameworks}, ACCO's baseline toolchain, DeFiNES \cite{mei2023defines}, is an advanced DF auto-scheduler with silicon verification.
Therefore, to verify the efficacy of ACCO's upgraded DSE, this case study compares the Pareto-optimal scheduling results between ACCO and DeFiNES.

\begin{figure}[!]
\centerline{\includegraphics[width=\linewidth]{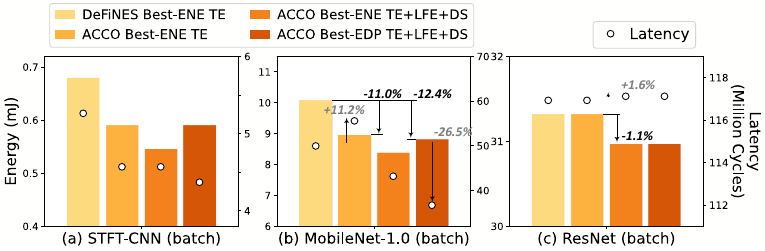}}
\caption{The SotA comparison of Pareto-optimal scheduling results between ACCO and  DeFiNES \cite{mei2023defines} on \textbf{TPU-Edge}. Three \textbf{batch-mode} workloads are considered: \textbf{a)} STFT-CNN, \textbf{b)} MobileNet-1.0, \textbf{c)} ResNet.
}
\label{fig:batch_sota}
\end{figure}

We conduct experiments across three batch-mode workloads (STFT-CNN, MobileNet-1.0, and ResNet) on TPU-Edge.
Fig. \ref{fig:batch_sota} \mvnok{summarizes the results across} different workloads. 
For STFT-CNN and MobileNet-1.0, ACCO's updated TE, LFE, and DS strategies show step-by-step advances from DeFiNES' optimums. 
Considering of the best EDP, ACCO achieves 22\% and 16.2\% in latency and energy reduction for the two workloads on average. 
Note that for MobileNet-1.0, the trend is a bit scattered, showing that ACCO discovers new \mv{Pareto}-optimal corners untouched by DeFiNES before.

On the contrary, the outcome of ResNet's DSE are almost identical across the two toolchains, \mvnok{stemming from the large weight volume of ResNet}. 
From Layer-18, each single layer's weights in ResNet exceeds the on-chip memory capacity \mvnok{of the modeled hardware accelerators}. 
Hence, even if fused together by ACCO's TE+LFE+DS strategies, loading these layers' weights during depth-first execution is always the critical bottleneck \mvnok{dominating energy and latency efficiency.} 

\subsection{Case Study 4: SotA Comparison \mv{for} Real-time \mv{Execution}} \label{section:experiments_case_study_4}

\mvnok{Finally, we benchmark ACCO against SotA published CNN scheduling approaches for spatio-temporal real-time input updates in Table \ref{tab:workload_parameter_per_mode}}. 
We explore real-time-mode STFT-CNN and MobileNet-0.5 workloads, whose DNN Weights are small enough for the target hardware's on-chip GB. 

Three SotA \mvnok{approaches are used} 
for comparison.
As specified in Table \ref{tab:sota_tool_features}, DNNVM \cite{xing2019dnnvm} and DeFiNES \cite{mei2023defines} \mv{are} designed for fully-spatial DNN optimization and cannot resolve the time-dimension causal relationship between spatio-temporal workloads.
\camready{
On the contrary, the TCN \cite{lea2016temporal} is a dedicated DNN for time-sequence reasoning. Its dilated causal structure is a corner case of ACCO when fusing all layers into one DF block, hence only spatially scheduled here. 
}

\camready{
During comparison, DNNVM and DeFiNES tune the inference execution based on original ST-CNN input frames (\textbf{Baseline} mode, Table \ref{tab:workload_parameter_per_mode}), while TCN and ACCO explore the equivalent computation that yields a single row of causal output from the temporally-flattened input frame (\textbf{Real-time} mode, Table \ref{tab:workload_parameter_per_mode}).
}

\begin{figure}[!]
\centerline{\includegraphics[width=\linewidth]{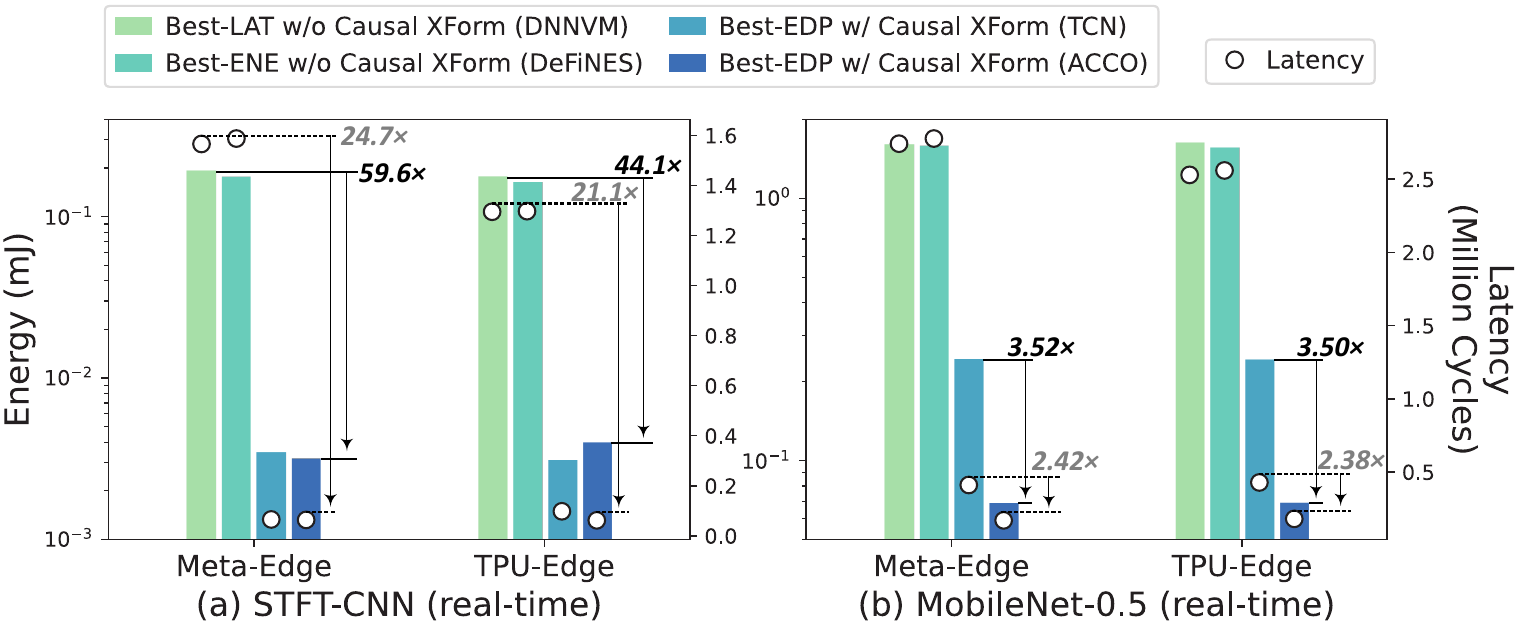}}
\caption{The SotA comparison of \textbf{Real-Time} mode overhead for STFT-CNN and MobileNet-0.5. Four SotA research are considered: \textbf{a)} \textbf{DNNVM} \cite{xing2019dnnvm} and \textbf{b)} \textbf{DeFiNES} \cite{mei2021zigzag} for the original spatio-temporal input; \textbf{c)} \textbf{TCN} \cite{lea2016temporal} and \textbf{d)} \textbf{ACCO} (this work) for the transformed causal input.
The target accelerator architectures are \textbf{Meta-Edge} and \textbf{TPU-Edge}.  
}
\label{fig:real-time_sota}
\end{figure}

From the results shown in Fig. \ref{fig:real-time_sota}, it is clear that ST-CNNs benefit much from the causal transformation. 
Even though weight reuse is suppressed in the causal form of the workload, the vast reduction in computation requirements surpasses the benefits of plain DF computation reduction on the original workload.
Putting together both advantages, ACCO DSE makes STFT-CNN 22.9$\times$ faster with 51.5$\times$ less energy consumption. 
On MobileNet-0.5, the efficiency gain is 15.1$\times$ and 23.2$\times$, respectively.
The significant efficiency boost compared to the \camready{spatially-scheduled TCN approach} (3.5$\times$ in energy and 2.4$\times$ in latency) on MobileNet-0.5 further confirms ACCO DSE's advantage in optimizing deep-and-narrow ST-CNN workloads. 

\begin{table}[t]
    \centering
    \caption{SotA ST-CNN optimizer Comparison}
    \label{tab:sota_tool_features}
    \renewcommand\arraystretch{1}
    \begin{tabular}{r|c|c|c}
    \toprule[0.22ex]
        & \makecell[c]{\textbf{Optimization}\\\textbf{Target}}& \makecell[c]{\textbf{Causal }\\\textbf{Awareness}} & \makecell[c]{\textbf{\camready{Schedulable}}\\\textbf{\camready{Dimension}}}\\ \Xhline{0.22ex}
         \textbf{DNNVM \cite{xing2019dnnvm}} & LAT & $\times$ & \camready{Spatial}\\ \hline
         \textbf{DeFiNES \cite{mei2023defines}} & LAT, ENE & $\times$ & \camready{Spatial} \\ \hline
         \textbf{TCN \cite{lea2016temporal}} & - & $\checkmark$ & \camready{Spatial} \\ \hline
         \textbf{ACCO (ours)} & LAT, ENE, EDP & $\checkmark$ & \camready{Spatial+Temporal} \\ 
    \bottomrule[0.22ex]
    \end{tabular}
\end{table}

\section{Conclusion}

This paper presents ACCO, an automated Causal CNN scheduling optimizer, which explores more efficient ST-CNN execution by \camreadyok{leveraging the computational reuse across time steps and CNN sliding windows.} 
It embeds a general \camreadyok{causal transformation} method to \mvnok{automatically} \mv{flatten} ST-CNNs 
and \camreadyok{identify} the previously-computed results.
Meanwhile, it extends the SotA DF scheduling toolchain to automatically optimize the transformed spatio-temporal workload \mvnok{across one or multiple frames}. 
In this way, for the causal-specific optimization, ACCO reaches 8.4$\times$ better EDP compared to the fixed \camreadyok{dilated-causal conversion}.
For the DF-only optimization, ACCO pushes forward the Pareto-optimal frontier by $\sim$20\% \mv{compared to} the SotA \mv{DF} DSE toolchain.
For the joint optimization of ST-CNNs, the best solutions from ACCO are 19$\times$ faster and 37$\times$ more energy-efficient than single spatial DF schemes on average.

\bibliographystyle{ieeetr}
\bibliography{bibli}






\end{document}